# Kitaev Spin Liquid Candidate Os$_x$Cl$_3$ Comprised of Honeycomb Nano-Domains

K. Kataoka[1], D. Hirai[1], T. Yajima[1], D. Nishio-Hamane[1], R. Ishii[1], K.-Y. Choi[2], D. Wulferding[3,4], P. Lemmens[3],
S. Kittaka[1,5], T. Sakakibara[1], H. Ishikawa[1], A. Matsuo[1], K. Kindo[1], and Z. Hiroi[1]

[1]*Institute for Solid State Physics, University of Tokyo, Kashiwa, Chiba 277-8581, Japan*
[2]*Department of Physics, Chung-Ang University, Seoul 06974, Republic of Korea*
[3]*Institute for Condensed Matter Physics, TU Braunschweig, 38106 Braunschweig, Germany*
[4]*Center for Correlated Electron Systems, Institute for Basic Science (IBS), Seoul 08826, Republic of Korea*
[5]*Department of Physics, Chuo University, Kasuga, Bunkyo-ku, Tokyo 112-8551, Japan*

An osmium chloride with the chemical formula of Os$_x$Cl$_3$ ($x$ = 0.81) was synthesized and its crystal structure and thermodynamic properties were investigated. Os$_x$Cl$_3$ crystallizes in a layered CdCl$_2$-type structure with the triangular lattice partially occupied by Os ions on average. However, on microscopic length scales, the triangular lattice is composed of nano-domains with a honeycomb arrangement of Os ions, as observed by electron microscopy and Raman scattering experiments. Magnetization and heat capacity measurements revealed an absence of magnetic long-range order down to 0.08 K, while a broad peak in heat capacity at 0.15 K may indicate a short-range order in the local honeycomb lattice. Os$_x$Cl$_3$ may exhibit certain aspects of the Kitaev spin liquid that are expected for a perfect honeycomb lattice of osmium trichloride.

## 1. Introduction

A quantum spin liquid is a quantum-mechanically entangled state in which the long-range magnetic order is absent at zero temperature even in the presence of strong interactions between the spins in the crystalline solid.[1] The state is typically realized in one-dimensional spin systems and only rarely observed in higher-dimensional ones. Recently, the ground state of the two-dimensional Kitaev model has been proven to be a quantum spin liquid.[2,3] This Kitaev spin liquid (KSL) state is induced by Ising-like, bond-dependent interactions between $S$ = 1/2 spins on a honeycomb lattice. Interestingly, the KSL is characterized by the novel fractional excitations of itinerant Majorana fermions and localized Z$_2$-fluxes.[2,4]

KSL states are predicted to occur in honeycomb compounds that features magnetic ions in a spin–orbit (SO) coupled $J_{eff}$ = 1/2 state.[5–9] The KSL candidates thus far studied have been low-spin $d^5$ compounds of α-RuCl$_3$, A$_2$IrO$_3$ (A = Li, Na or Cu) and related materials,[6,10–14] high-spin $d^7$ Co$^{2+}$ compounds,[15–17] and YbCl$_3$ and Na$_2$PrO$_3$ with the $f^1$/$f^{13}$ electron configurations.[18,19] Typically, the honeycomb lattices of magnetic ions in these compounds are embedded in two-dimensional layers composed of edge-sharing octahedra, where super-exchange interactions between the $J_{eff}$ = 1/2 spins become Kitaev interactions $K$. However, an additional Heisenberg interaction $J$ and off-diagonal interaction $Γ$ are non-negligible; as a result, the Hamiltonian of the $J$–$K$–$Γ$ model[20–22]

$$H_{ij} = -K^\gamma S_i^\gamma S_j^\gamma + J\mathbf{S_i} \cdot \mathbf{S_j} + \Gamma\left(S_i^\alpha S_j^\beta - S_i^\beta S_j^\alpha\right)$$

should be considered, where the indices {$α, β, γ$} = {$y, z, x$}, {$z, x, y$} and {$x, y, z$} represent for X-, Y- and Z-bonds between the sites i and j, respectively. Near the Kitaev limit (small $J$ and $Γ$), a spin liquid state is realized. However, in actual materials, magnetic orders tend to be induced by finite $J$ and $Γ$.

A$_2$IrO$_3$, which features a honeycomb lattice of Ir$^{4+}$ ions, has a $K$ value of 100–300 K[23–25] and exhibits a magnetic order below 10 K. H$_3$LiIr$_2$O$_6$, which features a similar structure, shows no magnetic order down to 0.05 K, although the experimental evidence of KSL has remained unclear, owing to quenched disorders.[14] In contrast, α-RuCl$_3$ seems a better candidate for KSL;[13,22,26–28] it features a honeycomb arrangement of Ru$^{3+}$ ions with the 4$d^5$ electron configurations and SO-coupled $J_{eff}$ = 1/2 moments; the honeycomb structure is slightly distorted at room temperature (space group $C2/m$), while the distortion is removed at low temperatures because of its structure transition to a trigonal structure ($R$–3).[29,30] The estimated values of $K$ are within 50–150 K,[23,31–33] and the compound exhibits a zigzag antiferromagnetic order at $T_N$ = 7 K, caused by the smaller non-Kitaev interactions.[28] Meanwhile, incoherent electronic excitations observed by Raman scattering[31] and inelastic neutron diffraction measurements[32] suggest the existence of the itinerant Majorana fermions;[3] this indicates that the ground state of α-RuCl$_3$ is located near a KSL state.

Further evidence of KSL states in α-RuCl$_3$ was obtained from thermal conductivity measurements in the presence of magnetic fields.[34] When the magnetic transition was suppressed through application of in-plane magnetic fields,[35] a half-integer quantum thermal Hall effect was observed; this reflects fractional Majorana excitations and seems to constitute direct evidence of a KSL. However, in the $J$–$K$–$\Gamma$ model, the survival of the KSL in the presence of a magnetic field has not yet been theoretically established. Hence, we require a new candidate compound exhibiting relatively large Kitaev interactions to study the KSL state in a zero magnetic field.

In this study, we focus on the $5d^5$ osmium trichloride. Its $J_{\text{eff}} = 1/2$ state is anticipated to be more stable than that of α-RuCl$_3$ because of its larger SO-coupling, which should enhance Kitaev interactions. Osmium trichloride was first synthesized in 1963.[36] Its powder X-ray diffraction (XRD) pattern was reported to resemble that of α-RuCl$_3$, and magnetic transitions were absent down to 78 K. Recently, McGuire et al. reported non-stoichiometric osmium chloride Os$_{0.55}$Cl$_2$,[37] which seems to be related to osmium trichloride. The crystal structure of this compound is based on a triangular lattice in which only 55% of the metal sites are occupied; this forms a 4×4×1 superlattice in the plane. The nominal valence state of the Os ions is 3.64, which is much larger than the value of 3 provided for osmium trichloride. No magnetic transition has been observed down to 0.4 K in magnetization and heat capacity measurements.

We attempt to synthesize osmium trichloride via various preparation methods, finally obtaining a single-phase sample by heating Os metal at 500 °C under Cl$_2$ gas in a quartz ampule. The obtained sample contains a large number of Os vacancies, and its chemical formula is Os$_x$Cl$_3$ ($x = 0.81$); this is similar to that of Os$_{0.55}$Cl$_2$ ($x = 0.83$). Although the global structure of Os$_x$Cl$_3$ also resembles that of Os$_{0.55}$Cl$_2$, the local structures are essentially different. In electron diffraction and transmission electron microscopy, we do not observe the 4×4×1 superlattice observed for Os$_{0.55}$Cl$_2$; instead, a weak and broad √3×√3×1 superlattice (corresponding to a honeycomb structure) is reported. The honeycomb structure exists locally in domains of approximately 2 nm. The presence of the honeycomb structure is also confirmed by Raman scattering experiments conducted on a single crystal. Magnetization measurements reveal that the Os ion has an effective paramagnetic moment of 1.36$\mu_B$ and a net ferromagnetic interaction of 8 K. Heat capacity measurements show that Os$_x$Cl$_3$ exhibits no magnetic transition down to 0.07 K; however, it may exhibit a short-range order below 0.15 K, which may occur within the local honeycomb structure of Os$_x$Cl$_3$. These results imply that stoichiometric osmium trichloride (with a perfect honeycomb structure of Os ions) represents a useful platform from which to investigate Kitaev physics.

## 2. Experiments
### 2.1. Synthesis and chemical analysis

Polycrystalline samples of Os$_x$Cl$_3$ were synthesized by heating Os metal under a Cl$_2$ atmosphere. We used 100 mg of pelletized Os metal (Alfa Aesar, 99.9%) and 0.4 ml of CCl$_4$ (Wako Junyaku, 99.5%) as a chlorine supplier; these were placed into quartz tubes of 10 mm inner diameter and approximately 200 mm length. After sealing under vacuum conditions (using liquid nitrogen to prevent CCl$_4$ evaporation), the quartz tubes with the pellet at one side were heated at 500 °C for 100 h. Provided the CCl$_4$ decomposed completely, the molar ratio would become Os:Cl = 1:23, yielding a partial pressure for Cl$_2$ of approximately 6 atm at 500 °C. This pressure was smaller than the pressure resistance of our silica tubes, which was approximately 11 atm. After heating, a black polycrystalline sample and numerous small, single crystals were obtained from the other side of the tubes. The obtained crystals possessed a plate-like shape with a maximum diameter of 100 μm and a thickness of less than 10 μm. The other products left in the tubes were a light green transparent liquid (possibly CCl$_4$ with dissolved Cl$_2$) and transparent C$_6$Cl$_6$ single crystals. These byproducts were removed by washing with hexane. When the quantity of CCl$_4$ was doubled, we obtained orthorhombic OsCl$_4$ instead of Os$_x$Cl$_3$[38].

The chemical composition of polycrystalline Os$_x$Cl$_3$ was determined by Rutherford backscattering (RBS) measurements. This technique is more reliable for performing chemical analyses than inductively coupled plasma optical emission spectroscopy, owing to the high volatility of OsO$_4$ and HCl in the solution. In the RBS measurements, $^4$He$^{2+}$ ions were accelerated to 2300 keV and made to collide into the sample; the composition was determined by measuring the energy dependence of the numbers of scattered ions. The measured molar ratio was Os:Cl = 0.81(2):3, which was close to the value of 0.83:3 reported for Os$_{0.55}$Cl$_2$.[37]

### 2.2. Structural analyses

The crystal structure of Os$_x$Cl$_3$ was determined via powder XRD analysis. XRD patterns were collected at room temperature using synchrotron X-ray radiation at an energy of 20 keV, at the beamline TPS09A of the National Synchrotron Radiation Research Center in Taiwan. To suppress the effect of the preferential orientation of crystallites, the experiment was performed in the transmission geometry, with the sample loaded into a quartz capillary that rotated during measurement. To obtain information on the local structure, transmission electron microscopy (TEM) and electron diffraction (ED) experiments were carried out in a JEOL JEM–2010F with an acceleration voltage of 200 kV. A crushed powder sample was used for TEM measurements.

### 2.3. Physical characterization

Raman scattering experiments were performed using a Horiba LabRam HR800 spectrometer equipped with a 532 nm Nd:YAG laser and a liquid-nitrogen-cooled charge-coupled device (Dilor CCD–One 3000). The laser beam was focused onto a spot (diameter: ~5 μm) on a single crystal sample using a microscope objective lens with 50 × magnification, and the scattered light was detected using a holographic ultra-low-frequency notch filter set. Temperature-dependent experiments were carried out between 5 and 300 K in a continuous He-flow cryostat (CryoVac KONTI Micro).

Magnetization measurements were carried out using an MPMS–3 SQUID magnetometer (Quantum Design) at 2–350 K, as well as a homemade capacitive Faraday magnetometer equipped with a dilution refrigerator effective down to 0.07 K.[39] Magnetization was measured up to 60 T at 1.4 K via an induction method, which used a pickup coil in pulsed magnetic fields generated by a multilayer pulse magnet. Heat-capacity experiments were performed using homemade equipment with a dilution refrigerator operating at 0.08–2 K under magnetic fields of up to 7 T. Polycrystalline samples were used for magnetization and heat-capacity measurements, because the single crystals were too small to provide reliable data.

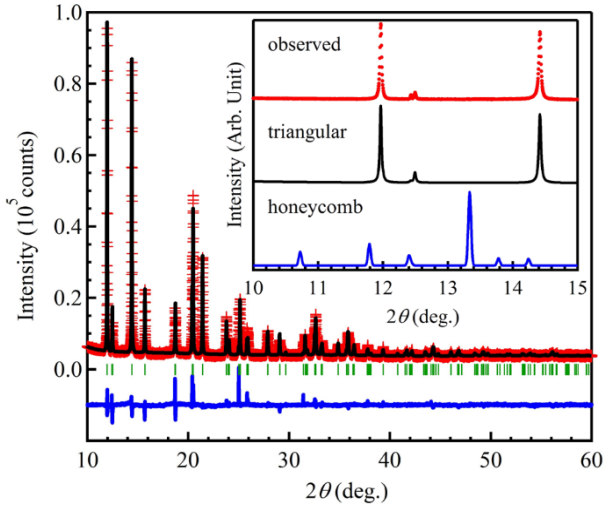

Fig. 1. Powder synchrotron X-ray diffraction patterns of Os$_x$Cl$_3$ at room temperature. The red crosses and solid black lines represent the observed data and Rietveld fit, respectively. The difference between them is indicated by the bottom blue line and the expected peak positions are shown as vertical green marks. The inset expands the low-angle region of the observed profile (top), which is compared with simulated profiles for the triangular (middle) and honeycomb structures (bottom).

Table 1. Structural parameters of Os$_x$Cl$_3$, determined by the Rietveld fit to the powder XRD data in Fig. 1; space group $R\text{–}3m$, $a = 3.48477(3)$ Å and $c = 17.1715(3)$ Å. $g$ and $U_{iso}$ are the occupancy and isotropic atomic displacement parameters, respectively.

| Atom | Site | $g$ | $x$ | $y$ | $z$ | $100 U_{iso}$(Å$^2$) |
|---|---|---|---|---|---|---|
| Os | 3$a$ | 0.559(2) | 0 | 0 | 0 | 1.004(18) |
| Cl | 6$c$ | 1 | 0 | 0 | 0.25561(12) | 2.06(5) |

## 3. Results
### 3.1 Crystal structure

The average crystal structure of Os$_x$Cl$_3$ was determined using powder XRD analysis (Fig. 1). All the observed diffraction peaks can be assigned to the indices of an $R$ lattice with the lattice constants $a = 3.48477(3)$ Å and $c = 17.1715(3)$ Å; this $R$ lattice suggests the existence of a threefold rotation symmetry. The value of $c$ is comparable to the tripled interlayer distance of 17.08 Å observed for α-RuCl$_3$ with the space group $C2/m$, suggesting a similar layered structure for Os$_x$Cl$_3$. However, the lattice constant $a$ differs from the $a = 5.96$ Å of α-RuCl$_3$, which assumes the space group $P3_112$; its value here is $1/\sqrt{3}$ times smaller and close to its typical value for the CdCl$_2$-type layered structure with the space group $R\text{–}3m$. It should be noted that the honeycomb structure in α-RuCl$_3$ is regarded as a $\sqrt{3}a \times \sqrt{3}a$ superlattice, obtained by systematically removing one-third of the metal ions from the triangular lattice of the CdCl$_2$ structure. The composition of $x = 0.81(2)$ indicates the presence of many vacancies in the Os site.

Rietveld refinement was performed, assuming a CdCl$_2$-type structure (Fig. 1). The refinements resulted in an $R_{wp}$ of 5.76%, which is fairly low; this is possibly due to the presence of stacking faults either inherent in the material or generated through grinding the powder, or it may be due to the short-range order of the Os ions in the plane, as discussed below. The obtained atomic parameters are listed in Table 1; Os and Cl occupy the 3$a$ and 6$c$ sites, respectively. The Os site is occupied by 0.559(3), providing a ratio of $x = 0.839(3)$, which is close to the chemical composition of $x = 0.81(2)$ found from the RBS measurements. Considering the insufficient quality of the Rietveld analysis, we take the latter value for Os$_x$Cl$_3$.

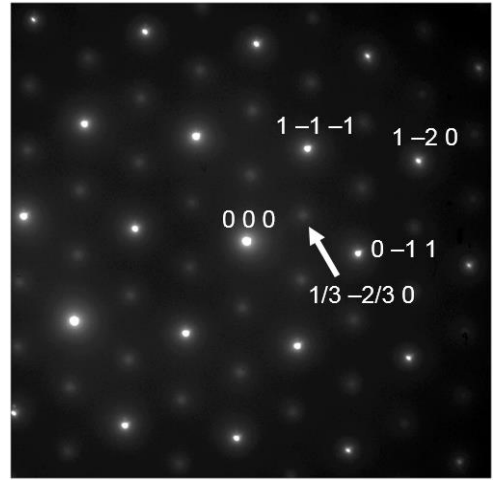

Fig. 2. Electron diffraction pattern of Os$_x$Cl$_3$ with the incident electron beam along the [211] direction which is inclined at 19° from the $c$ axis. The indices of diffraction spots assume a CdCl$_2$-type unit cell with a triangular arrangement. The intense spots are fundamental reflections, whereas the weak and broad spots such as marked by the arrow, (1/3, –2/3, 0), are superlattice reflections based on the in-plane $\sqrt{3}a \times \sqrt{3}a$ unit corresponding to a honeycomb structure.

TEM observations were made to gain insight into the local structure of Os$_x$Cl$_3$. An ED pattern formed with an incident electron beam along the [211] zone axis is shown in Fig. 2; this incident direction is inclined at 19° from the $c$ axis. Intense, sharp fundamental reflections and weak, broad superlattice reflections were observed. The six intense spots around the direct-beam spot, (0 0 0), correspond to a lattice spacing of $d \sim 3.0$ Å and are labeled with {0 −1 1} and {1 −1 −1} indices based on the $CdCl_2$ structure; meanwhile, the six inner spots correspond to $d \sim 5.1$ Å. The typical spot (marked by the arrow) can be assigned to the index (1/3 −2/3 0), which divides the (1 −2 0) vector into three and corresponds to a $\sqrt{3} \times \sqrt{3} \times 1$ superlattice. The broadening of these reflections indicates either that the superlattice exists only locally or that many disorders are present.

The origin of the superlattice reflections was identified by high-resolution lattice images. Figure 3(a) shows a typical lattice image in which the electrostatic potential of a triangular array of metal atoms is projected. It should be noted that a significant modulation in the contrast is observed on the nanometer scale. The Fourier transform of this lattice image [shown in Fig. 3(b)] captures the features of the ED pattern (shown in Fig. 2); here, the outer and inner sets of the six spots surrounding the origin correspond to the $a \times a$ fundamental and $\sqrt{3}a \times \sqrt{3}a$ superlattice reflections, respectively. The inverse Fourier-transformed images are shown in Figs. 3(c) and (d), respectively. The former shows a triangular arrangement of $a \times a$, whereas the latter shows a larger arrangement of $\sqrt{3}a \times \sqrt{3}a$. Notably, the $\sqrt{3}a \times \sqrt{3}a$ arrangement is significantly modulated on a scale of only a few nanometers. A regular pattern of bright spots of $\sqrt{3}a \times \sqrt{3}a$ appears in a domain with a diameter of ~2 nm, and the region between the nearby domains has an obscured contrast, indicating a disordered arrangement of metal atoms between domains. Furthermore, the triangular pattern in Fig. 3(c) exhibits a weak modulation in contrast; this suggests a modulation in the triangular lattice, caused by the formation of the honeycomb domains. Concerning the size of the domains, the 2 nm diameter observed may be a minimum estimate, because the image should give a projection along the $c$ axis; the actual in-plane domain sizes could be larger if little structural correlation exists along this axis.

Figure 4 presents schematic pictures of the arrangements of Os ions in a layer of $Os_xCl_3$. The average structure is a triangular lattice with 56% occupancy of the Os site, as shown in Fig. 4(a). The ideal honeycomb structure [shown in Fig. 4(c)] corresponds to a $\sqrt{3}a \times \sqrt{3}a$ superlattice. Our TEM observations indicate that the distribution of vacancies is not random but instead forms a honeycomb structure with domains of approximately 2 nm in diameter, as presented in Fig. 4(b). Thus, we conclude that $Os_xCl_3$ forms a depleted triangular lattice with nanometer-scale honeycomb domains.

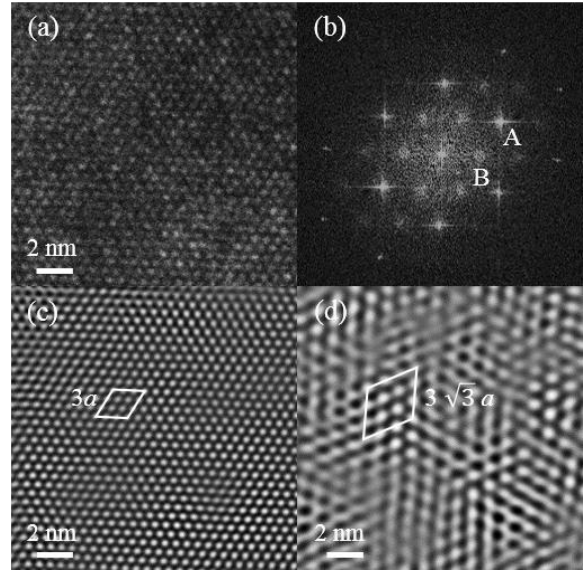

Fig. 3. (a) Real-space transmission electron microscopy image, approximately taken at the [001] zone axis, showing a projection of the electrostatic potential along the $c$ axis. (b) Fourier transform image of (a), indicating intense and weak spots as A and B, respectively. (c) Inverse Fourier transform image from the set of A-type spots in (b). (d) Inverse Fourier transform image from the set of B-type spots in (b). The rhomboid in (c) outlines a $3a \times 3a$ cell for the triangular structure, and that in (d) outlines a $3\sqrt{3}a \times 3\sqrt{3}a$ cell of the honeycomb structure. The bright or dark spots in (d) correspond to the hexagonal holes of the honeycomb lattice. The honeycomb structure occurs only locally, on the scale of a few nanometers.

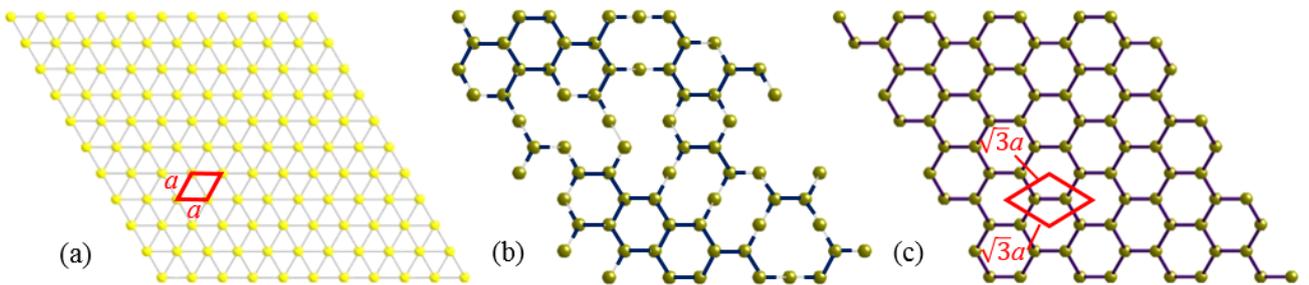

Fig. 4. Arrangements of Os atoms in a layer of $Os_xCl_3$. (a) Triangular structure for $Os_{1.5}Cl_3$. The $a \times a$ unit cell is depicted by the rhomboid. (b) Schematic representation of a short-range honeycomb order with nm-size domains, which must be realized in our sample; 44% of the Os atoms have been quasi-periodically removed from (a), to generate segmented honeycomb lattices. (c) Ideal honeycomb order of the $\sqrt{3}a \times \sqrt{3}a$ superlattice marked by the rhomboid for $OsCl_3$.

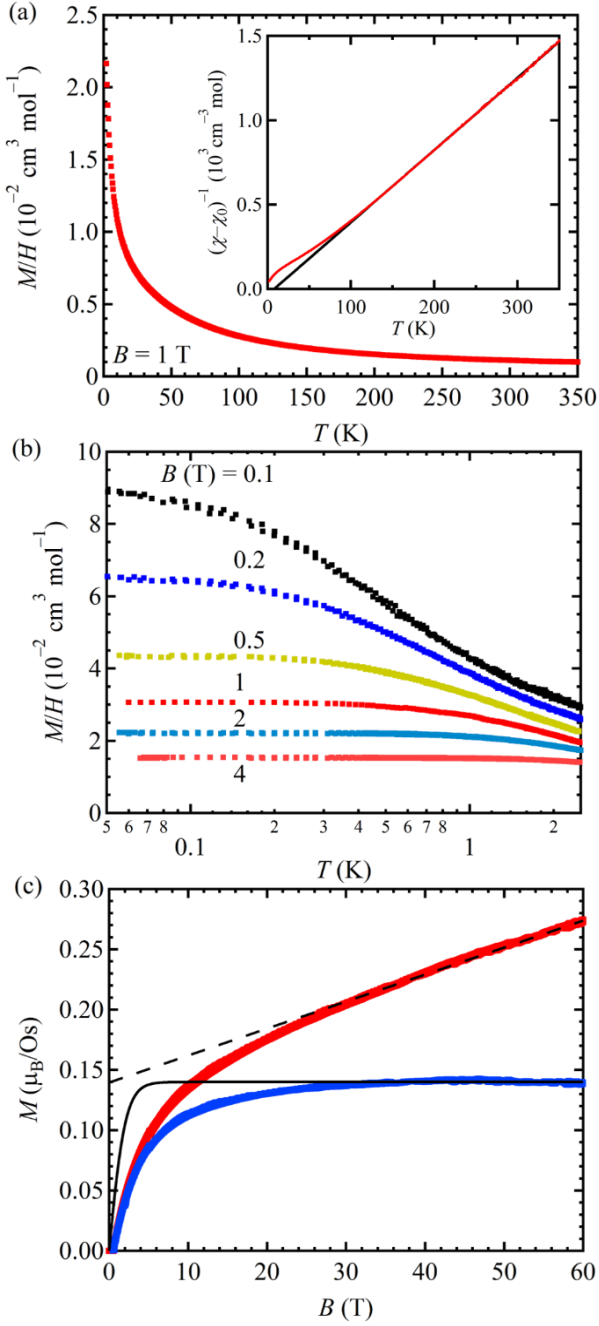

Fig. 5. (a) Temperature dependence of the magnetic susceptibility of a polycrystalline sample of Os$_x$Cl$_3$, measured in a magnetic field of 1 T down to 2 K. The inset depicts the inverse susceptibility with a Curie−Weiss fit shown by the solid black line; the temperature-independent term $\chi_0 = 3.10(3) \times 10^{-4}$ cm$^3$ mol$^{-1}$ has been subtracted. (b) Field evolution of low-temperature magnetic susceptibilities between 0.07 and 2.5 K, measured at $B$ = 0.1–4 T. (c) Field dependence of magnetization (red line) at 1.4 K. The broken black line on the data points represents a linear fit at 30–60 T. A nonlinear component (blue line) remaining after the subtraction of the linear component saturates more gradually than predicted by the Brillouin curve calculated for non-interacting spins.

### 3.2 Magnetic properties

The temperature dependence of the magnetic susceptibility of Os$_x$Cl$_3$ is shown in Fig. 5(a). The susceptibility tends to increase smoothly upon cooling, and it does not show any anomaly indicative of a magnetic order down to 2 K. A Curie–Weiss fitting was performed in the range of 150–350 K for the equation $\chi = C/(T − \theta_{cw}) + \chi_0$, where $C$ denotes the Curie constant, $\theta_{cw}$ is the Weiss temperature, and $\chi_0$ is a temperature-independent term. The parameters obtained from the fitting are $C$ = 0.233(1) cm$^3$ K mol-Os$^{-1}$, $\theta_{cw}$ = 8.0(6) K, and $\chi_0$ = 3.10(3)×10$^{-4}$ cm$^3$ mol$^{-1}$. The effective moment per osmium ion was calculated to be 1.36$\mu_B$. Provided $J_{eff}$ = 1/2 and 0 for Os$^{3+}$ and Os$^{4+}$ ions, their effective moments are 1.73$\mu_B$ and 0, respectively. Considering the chemical molar ratio of 3:7 between them, the expected average moment would be 0.95$\mu_B$, much smaller than the observed value. Thus, the Os$^{3+}$ ions should have an enhanced moment, as reported for other $d^5$ compounds of Ru$^{3+}$, Rh$^{4+}$, and Ir$^{4+}$ ions,[10,26,40] and/or the Os$^{4+}$ ion is not perfectly nonmagnetic but has a small magnetic moment, as suggested for other $d^4$ states of Os$^{4+}$ or Ir$^{5+}$.[41,42]

The observed positive Weiss temperature of 8 K indicates a net ferromagnetic interaction between the Os ions. However, it should be noted that, below 100 K, the susceptibility shown in Fig. 5(a) begins to deviate from the Curie–Weiss curve toward smaller values, which suggests the presence of comparable ferromagnetic Kitaev interactions. Thus, we expect that competition between ferromagnetic Kitaev interactions (which are effectively as large as 100 K) and smaller antiferromagnetic non-Kitaev interactions reduces the absolute value of the Weiss temperature.

No anomaly suggestive of a magnetic transition was detected at 0.07–2 K in the low-$T$ magnetic susceptibility of Fig. 5(b). Upon cooling, the susceptibility gradually increases and saturates below 0.2 K in low fields, whereas the temperature dependence becomes weak with the increasing magnetic field. Figure 5(c) shows the field dependence of magnetization at 1.4 K. It features two components: a linear term increasing up to 60 T and a non-linear term that saturates at low magnetic regions. The former is a response from majority spins, and the latter may come from the fewer interacting spins at crystalline defects. From the intercept of the linear term, the contribution of these defect spins is estimated to be 0.14$\mu_B$ per Os. Compared with the Brillouin curve [represented by the solid black line in Fig. 3(c)], the non-linear component saturates in a larger field, suggesting that the defect spins are not simple orphan spins but instead interact with neighboring spins with reduced magnetic interactions.

### 3.3 Heat capacity

The temperature dependence of $C/T$ for Os$_x$Cl$_3$ is plotted in Fig. 6(a). For a zero magnetic field, a broad peak is observed at approximately 0.15 K. The position and intensity of the peak remains intact at 0.1 and 0.2 T, whereas it shifts towards higher temperatures and becomes broad under a further increase of magnetic fields. The same set of data is plotted as a function of $T/B$ (in the logarithmic scale) in Fig. 6(b), where the peak moves to the left up to 0.5 T and remains above 0.5 T. Thus, the peak temperature is proportional to the magnetic field only above 0.5 T, which is characteristic of a Schottky type heat capacity featuring a doubly degenerate energy level for a free spin that it linearly split by the Zeeman energy. The origin of this Schottky type heat capacity could be weakly coupled spins generated by Os ions near the defects of the honeycomb lattice.

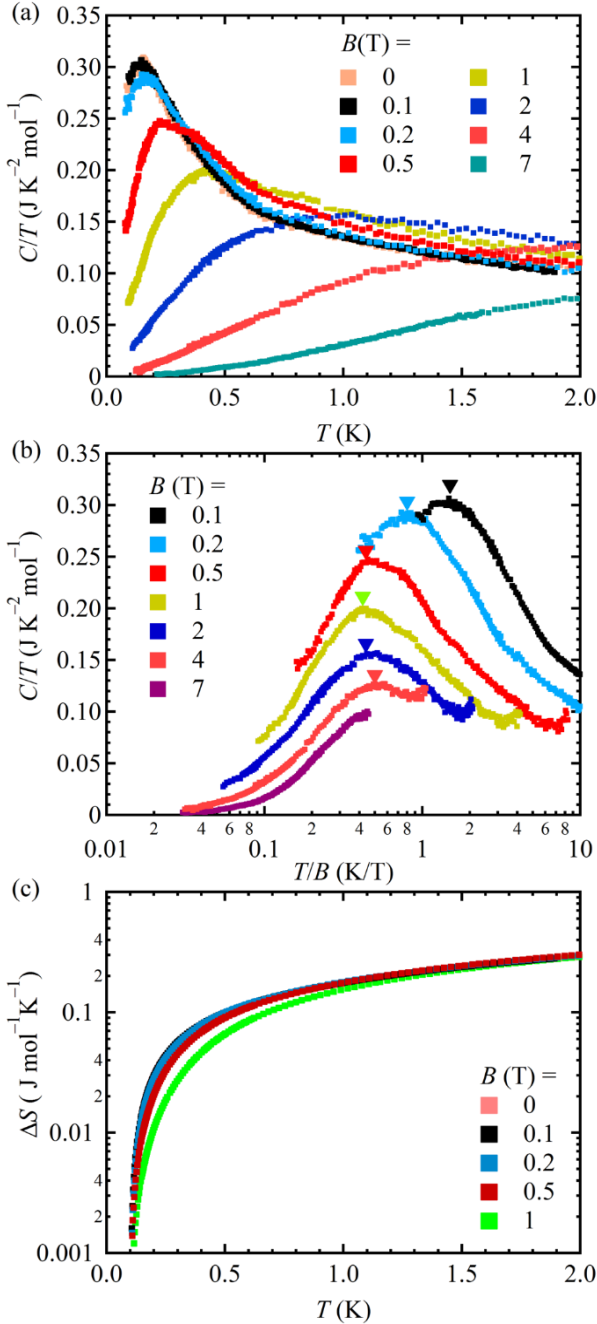

Fig. 6. (a) Temperature dependence of heat capacity divided by temperature under magnetic fields at $T = 0.08$–2 K, measured using a dilution refrigerator. (b) $C/T$ plotted as a function of $T/B$. The triangles mark the top of the peaks. (c) Variations in entropy from 0.1 K to 2 K, obtained from the $C/T$ data.

However, the peak at 0.15 K must be caused by the Os ions within the regular honeycomb structure, because it appears under the zero-field condition and is insensitive to small magnetic fields. With an increasing magnetic field, this peak must be suppressed and eventually hidden under the Schottky contribution. In fact, the data at $B = 0.5$ T show a double peak that may represent the sum of the two contributions. It is likely that magnetic freezing sets in at 0.15 K. Because of the nm-size honeycomb domains, a long-range order cannot be realized, although a short-range order occurs. The absence of a corresponding anomaly in the magnetic susceptibility of Fig. 5(b) may be due to the weak magnetic transition.

The variation in entropy is estimated by integrating $C/T$ between 0.1 and 2 K [Fig. 6(c)]; below 2 K, the small lattice contribution to heat capacity is negligible. The released entropy is 0.29 J K$^{-1}$ mol$^{-1}$, which corresponds to 5.1% of $R\ln 2$. Thus, most of the magnetic entropy should have been released at higher temperatures, and the defect-spin contribution should dominate for this low-temperature heat capacity. The Kitaev model predicts that the entropy is released successively at two well-separated temperature scales;[4] this is indeed observed for α-RuCl$_3$. Unfortunately, we were not successful in evaluating the magnetic entropy of Os$_x$Cl$_3$ over a wide temperature range, owing to the difficulty in subtracting the lattice contribution. Although we synthesized IrCl$_3$ with a honeycomb structure and filled the $t_{2g}$ orbitals to approximate the lattice heat capacity, the heat capacity of the two compounds did not match at high temperatures, exhibiting a significant difference in the lattice contribution[38]. For further investigation, we require an osmium chloride with fewer defects.

*3.4 Raman spectroscopy*

Raman scattering experiments were performed on a single crystal of Os$_x$Cl$_3$, to extract information on its crystal structure and its low-energy, electronic excitations. Figure 7(a) compares the Raman spectra of Os$_x$Cl$_3$ and α-RuCl$_3$ at 300 K. It is evident that the number of phonon modes in Os$_x$Cl$_3$ exceeds that in α-RuCl$_3$ (space group $C2/m$) at ambient pressure.[43] Meanwhile, we noted a resemblance between Os$_x$Cl$_3$ at ambient pressure and α-RuCl$_3$ at 5 GPa (space group $C2$).[44] This suggests a similarity in the local crystal environments between the OsCl$_6$ and RuCl$_6$ octahedra at high pressure. For α-RuCl$_3$, the dimerization of neighboring Ru ions is induced above a critical pressure of 1.7 GPa, leading to a splitting of phonon modes.[44] It is possible that the local environment around the Os ions in the disordered honeycomb net of Os$_x$Cl$_3$ resembles that of the high-pressure structure of α-RuCl$_3$.

Next, we attempted to assign the observed phonon modes in Os$_x$Cl$_3$ to their irreducible representations, based on the $C2/m$ space group; however, we cannot exclude the possibility that the local crystal symmetry is actually lower. From factor group analysis, we expect the Raman-active modes $\Gamma_{\text{Raman}} = 6A_g$ (*aa, bb, cc, ac*) + $6B_g$ (*ba, bc*). The $A_g$ and $B_g$ modes might be nearly degenerate, as reported for CrCl$_3$.[43] The 8.7 meV mode corresponds to the twisting motion of the Os–Cl–Os–Cl plane, and the 11.2 and 17.4 meV modes correspond to the in-plane relative Os movements. Furthermore, the 38.4 meV mode to the Os–Cl–Os–Cl shearing and the 45.9 meV mode with the breathing mode are assigned to the Os–Cl–Os–Cl ring structure.[44] The 48.4 and 55.8 meV modes are assigned to the symmetric and antisymmetric breathing modes between the upper and lower Cl layers, respectively,.[44] Notably, the 8.7, 17.4, and, 38.4 meV modes exhibit multiple-peak structures, reflecting the presence of various local Os environments.

In the Raman spectra at higher energies [Fig. 7(b)], a broader structure is observed, which consists of a well-defined peak $A_0$ at $195 \pm 5$ meV and two broader peaks $A_1$ and $A_2$ at $470 \pm 10$ and $650 \pm 50$ meV, respectively. The combined features of these excitations compare well to the electronic excitations reported for α-RuCl$_3$.[45] The $A_0$ peak was previously interpreted in terms of an SO exciton between the SO-coupled levels of $J_{eff} = 1/2$ and $J_{eff} = 3/2$, while the A1 and A2 peaks were assigned to transitions to SO-coupled $e_g$ states.[44] Recently, a resonant inelastic X-ray scattering study questioned these assignments.[46] Instead of the $A_0$ peak, the $A_1$ peak was identified as an SO exciton. Regarding the $A_2$ peak, two scenarios have been proposed; namely, a double-SO exciton or charge-transfer-type excitations from Cl 3$p$ to Ru 4$d$. Thus, the precise assignment of the three observed peaks is far from being clear for Os$_x$Cl$_3$ and lies beyond the scope of the present work. Nevertheless, we can estimate the lower value of the SO coupling in Os$_x$Cl$_3$. Assuming that the $A_0$ peak energy is related to the SO constant $\lambda \sim (2/3) \cdot A_0$, we obtain $\lambda = 130$ meV, which is larger than the $\lambda = 96$ meV found for α-RuCl$_3$.[45]

In addition to the phonon modes, we observed a weakly structured background in the energy range 0–65 meV at all temperatures, as shown by the red- and blue-shaded areas at 300 and 5 K in Fig. 8(a). The broad continuum is not due to the phonon density of states, because its temperature dependence is not in accordance with the effect of lattice disorders. Instead, it is ascribed to a continuum of magnetic excitations. However, compared with conventional spin-wave excitations, its $T$-dependence is rather subtle; its energy range is also unusually broad, without any clear finite onset energy. In contrast, it resembles the spectroscopic features of fractional excitations from a spin-liquid state.[31,47–49] A detailed study of its thermal evolution can help distinguish between conventional (bosonic) and fractional (fermionic) excitations. This approach has been successfully applied to the related Kitaev compounds α-RuCl$_3$, β-Li$_2$IrO$_3$, and γ-Li$_2$IrO$_3$.[31,43,48,50] In Fig. 8(b), we plot the $T$-dependence of the intensity of the magnetic continuum, which was obtained by integrating the magnetic spectra over the full energy range. We fit the intensity to the sum of a single bosonic scattering process [$(n + 1)$ with $n = 1/(\exp(\beta E_B) - 1)$] and a two-fermionic scattering process [$(1 - f)^2$ with $f = 1/(1 + \exp(\beta E_f))$]; this corresponds to the creation or annihilation process of a pair of Majorana fermions (see Ref. 50 for details). Our fit yields a characteristic two-fermionic energy scale $E_f = 5.4 \pm 0.5$ meV, which is remarkably close to the value reported for α-RuCl$_3$ ($E_f = 5.25$ meV).[43] This observation of the fermionic contribution supports the possibility that Os$_x$Cl$_3$ represents the KSL with a fractionalized excitation spectrum.

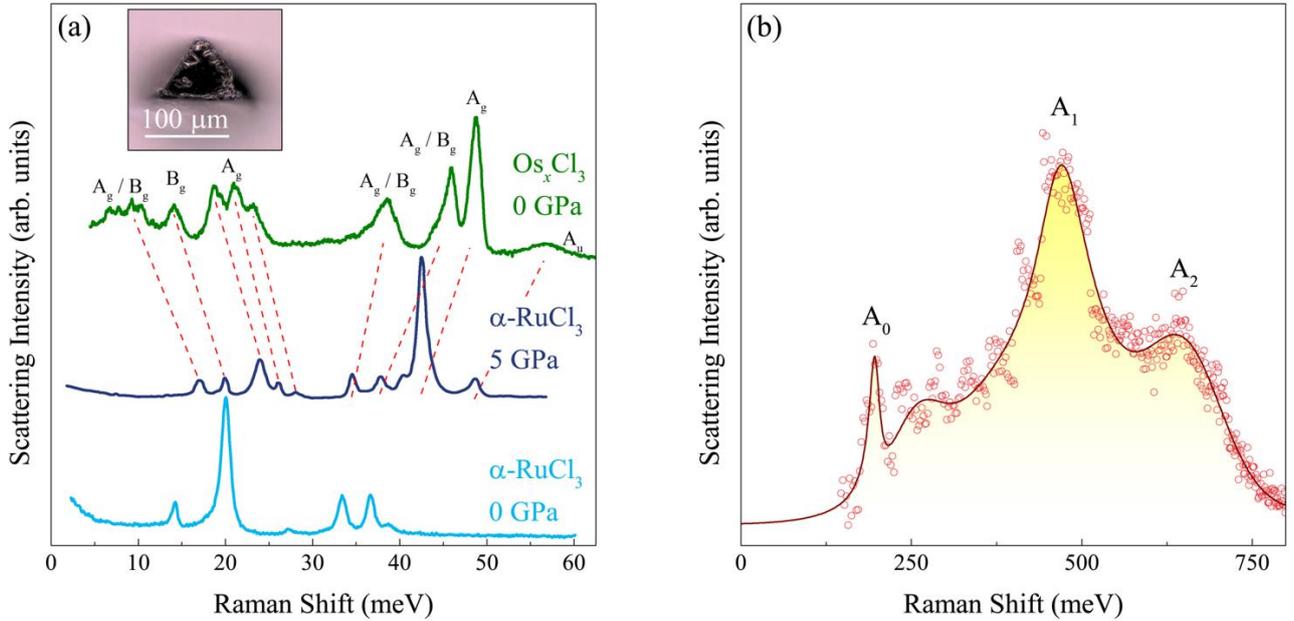

Fig. 7. (a) Raman spectra of Os$_x$Cl$_3$ at $T = 300$ K and ambient pressure. Those of α-RuCl$_3$ at 0 GPa and 5 GPa (adapted from Ref. 44) are shown for comparison. An optical microscope image of the Os$_x$Cl$_3$ single crystal used for measurements is shown in the inset. (b) Raman spectrum of Os$_x$Cl$_3$ in an extended energy range up to 750 meV. The curve is a fit to the data.

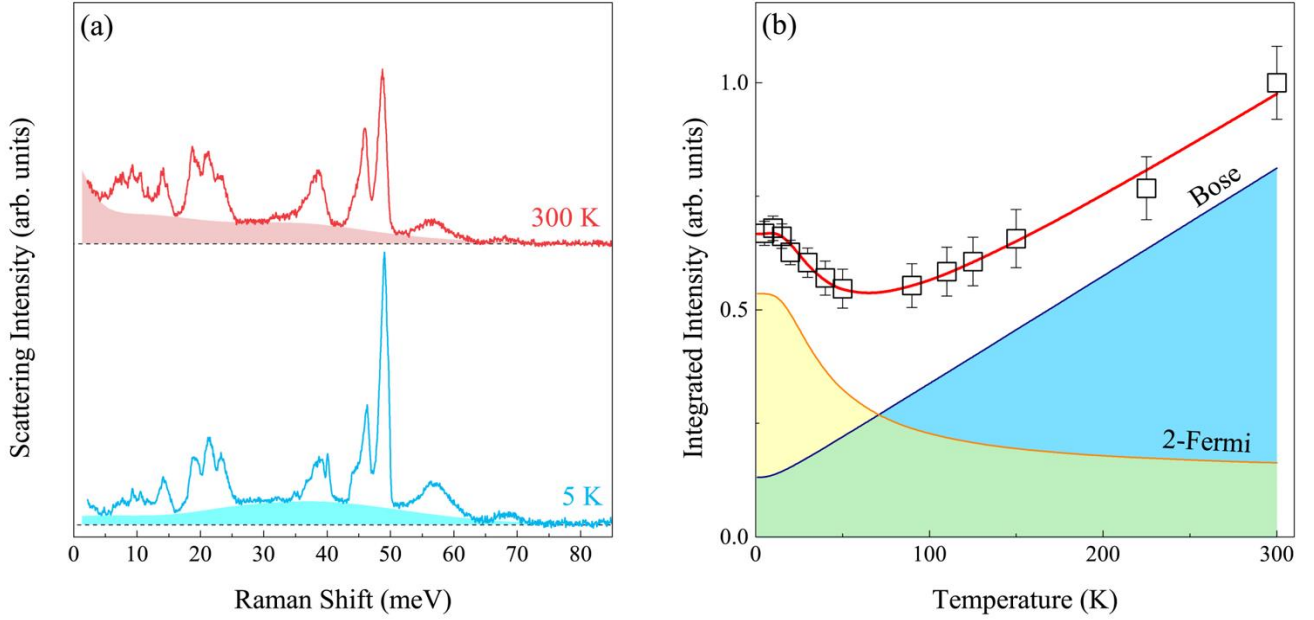

Fig. 8. (a) Temperature evolution of the continuum in the magnetic excitations at $T$ = 300 and 5 K (shaded in red and blue, respectively). (b) Temperature dependence of the integrated intensity of the magnetic continuum (open squares). The intensity is fitted to a sum of bosonic (blue) and two-fermionic (yellow) excitations.

## 4. Discussion

*4.1 Comparison to other Kitaev candidates*

In Table 2, the structural and magnetic properties of Os$_x$Cl$_3$ and the other Kitaev compounds are compared.[12,26,37,51] Typically, these feature $d^5$ electron configurations and honeycomb structures composed of edge-sharing octahedra. The nearest-neighbor metal–metal distance $d_{M-M}$ is comparable between Os$_x$Cl$_3$ and α-RuCl$_3$, while it is relatively smaller in iridium oxides. The difference arises from the different ionic radii of chloride and oxide ions. The deviation (from 90°) of the M–X–M bond angle $\phi$, which measures the distortion of an octahedron, is smallest for Os$_x$Cl$_3$. The relatively small $d_{M-M}$ and large distortion of iridium oxides enhance their non-Kitaev interactions. In fact, their Weiss temperatures are negative, indicating that non-Kitaev antiferromagnetic interactions dominate over the ferromagnetic Kitaev interactions. In contrast, the small positive Weiss temperature of Os$_x$Cl$_3$ suggests the presence of ferromagnetic Kitaev interactions that are comparatively stronger than their antiferromagnetic counterparts.

At 7 K, α-RuCl$_3$ exhibits the magnetic long-range order of a zigzag antiferromagnetic state.[26] For α-A$_2$IrO$_3$ (A = Li or Na), antiferromagnetic orders appear at 15 K;[10,11] however, for Cu$_2$IrO$_3$, $T_N$ = 2.7 K, which is lower than that of other compounds, in spite of their comparable Weiss temperatures; this is considered to be a result of the randomness associated with site mixing between the Ir and Cu ions.[12] In contrast, the magnetic order is suppressed to a lower temperature in Os$_x$Cl$_3$. Provided that the anomaly in heat capacity observed at 0.15 K is an indicator of a magnetic order (which may be broadened by the structural disorder and formation of nano-domains), the large reduction in transition temperature suggests that the ground state of Os$_x$Cl$_3$ is located close to a KSL. One possible explanation for the suppression of the magnetic order in Os$_x$Cl$_3$ is the relatively small non-Kitaev interactions. One source of non-Kitaev interactions is the distortion of the octahedra;[6,21,52] the magnitude of the non-Kitaev interactions depends on $\Delta/\lambda$, where $\Delta$ is the crystal field distortion. As described above, the distortion of octahedra in Os$_x$Cl$_3$ is smaller than that observed in α-RuCl$_3$, and the $\lambda$ = 130 meV of Os$_x$Cl$_3$ exceeds the $\lambda$ = 96 meV of the latter. Thus, $J_{eff}$ = 1/2 is stabilized with less non-Kitaev interactions, because of the larger $d_{M-M}$ and the $\phi$ values closer to 90°.

Table 2. Metal–metal distance ($d_{M-M}$), M–X–M bond angle ($\phi$), Curie–Weiss temperature ($\theta_{cw}$) and Néel temperature ($T_N$) in Os$_x$Cl$_3$, the other Kitaev candidates of α-RuCl$_3$ and A$_2$IrO$_3$ (A = Li, Na or Cu) and related compound Os$_{0.55}$Cl$_2$. The $d_{M-M}$ and $\phi$ are based on the crystal structures at 300 K. Upper and lower values of $\theta_{cw}$ are obtained from anisotropic susceptibilities $\chi_c$ and $\chi_{ab}$, respectively.

| Compound | $d_{M-M}$(Å) | $\phi$(°) | $\theta_{cw}$(K) | $T_N$(K) | Ref. |
|---|---|---|---|---|---|
| Os$_{0.81}$Cl$_3$ | 3.485 | 92.1 | 8 | (0.15) | This work |
| α-RuCl$_3$ | 3.449 | 93.6 | −145 / 68 | 7 | 26 |
| α-Li$_2$IrO$_3$ | 2.91 | 94.7 | −33 | 15 | 52 |
| α-Na$_2$IrO$_3$ | 3.12 | 98.0 | −126 | 15 | 52 |
| Cu$_2$IrO$_3$ | 3.07 | 94.6 | −110 | 2.7 | 12 |
| Os$_{0.55}$Cl$_2$ | 3.487 | 92.6 | −53.8 / 0.3 | none | 37 |

However, the influence of structural disorder on the reduction of $T_N$ must be considered in Os$_x$Cl$_3$. It is known that randomness induces spin-glass states or spin liquid-like states in Kitaev compounds.[53–55] To clarify this issue, an ideal OsCl$_3$ with a perfect honeycomb lattice is required.

Osmium trichlorides will provide us with a good platform to investigate Kitaev physics once a better sample with fewer structural defects has been obtained.

*4.2 Comparison to Os$_{0.55}$Cl$_2$*

Finally, we compared Os$_x$Cl$_3$ with the Os$_{0.55}$Cl$_2$ reported in Ref. 37. The chemical composition of Os:Cl = 0.81(2):3, obtained by RBS analysis, is almost identical to the value Os:Cl = 0.83(3):3 obtained for Os$_{0.55}$Cl$_2$. On average, both compounds have CdCl$_2$-type triangular structures. The lattice constants of $a$ = 3.48477(3) Å and $c$ = 17.1715(3) Å for Os$_x$Cl$_3$ are comparable to $a$ = 3.4874(6) Å and $c$ = 17.159(5) Å for Os$_{0.55}$Cl$_2$. Namely, their Os–Os distances and Os–Cl–Os bond angles are similar (Table 2). Meanwhile, the local arrangements of defects differ distinctively. The √3×√3×1 superlattice (corresponding to the honeycomb structure) is observed in Os$_x$Cl$_3$, whereas this type of superlattice remains subdominant in Os$_x$Cl$_3$. Instead, a well-defined 4×4×1 superstructure occurs in Os$_{0.55}$Cl$_2$; we did not observe such 4×4×1 superlattice reflections in our TEM experiments for Os$_x$Cl$_3$. Thus, in the nano-domains, the Os arrangements in the layer are of a non-honeycomb type in Os$_{0.55}$Cl$_2$ and of a honeycomb type in Os$_x$Cl$_3$. This difference is most likely due to differing conditions in sample preparation.

The temperature dependences of magnetization in these osmium chlorides are very similar to each other. No anomalies are observed down to 0.07 K and 0.4 K for Os$_x$Cl$_3$ and Os$_{0.55}$Cl$_2$, respectively. The Curie constant of 0.233(1) cm$^3$ K mol-Os$^{-1}$ for Os$_x$Cl$_3$ accords with a powder-averaged one of 0.24 cm$^3$ K mol-Os$^{-1}$ for Os$_{0.55}$Cl$_2$. In contrast, the Weiss constants differ significantly: $\theta_{cw}$ = 8 K for Os$_x$Cl$_3$ and −54 K ($H \parallel c$) and 0.3 K ($H \perp c$) for Os$_{0.55}$Cl$_2$, suggesting that a larger contribution of antiferromagnetic interactions is involved in the latter.

Heat capacity measurements were performed down to 0.08 K for Os$_x$Cl$_3$, which revealed broad anomalies at 0.15 K at low magnetic fields, and 0.4 K for Os$_{0.55}$Cl$_2$. Both chlorides show broad anomalies that vary with the applied field, which can be described as Schottky anomalies for Os$_x$Cl$_3$, while, for Os$_{0.55}$Cl$_2$, such anomalies may be due to the complicated energy levels associated with the local arrangement of Os$^{3+}$ and Os$^{4+}$ ions.[37]

Nevertheless, structural disorder has a strong influence on the physical properties of either compound, obscuring their intrinsic properties.

## 5. Conclusions

We synthesized a Kitaev candidate, osmium chloride Os$_x$Cl$_3$ ($x$ = 0.81), and we studied its structural and physical properties. This compound features nano-domains of the honeycomb lattice of Os ions. The magnetism was approximated as a $J_{eff}$ = 1/2 state, which is expected for a 5$d^5$ electron configuration in which the ferromagnetic Kitaev interactions dominate marginally over the antiferromagnetic Heisenberg interactions. Raman spectroscopy experiments revealed a large SO-coupling of $\lambda$ = 130 meV, larger than the $\lambda$ = 96 meV observed for α-RuCl$_3$. In addition, we observed a continuum of magnetic excitations induced by Kitaev interactions. An anomaly seemingly associated with long-range orders was not observed in the magnetization and heat capacity measurements down to 0.08 K. Instead, we observed a broad peak in heat capacity at 0.15 K, which may indicate a short-range magnetic order. The large suppression of magnetic order suggests that Os$_x$Cl$_3$ is closer to the intrinsic KSL phase than the other Kitaev candidates. To gain more insight into Kitaev physics, we must eliminate structural disorder and synthesize ideal OsCl$_3$ featuring the honeycomb structure.

**Acknowledgements**

We are grateful to Yukitoshi Motome for helpful discussions. The synchrotron radiation XRD experiments were performed at the beam line TPS09A of NSRRC in Taiwan (Project No. 2018-2-252-1).

This work was partially supported by KAKENHI (Grant No. JP19H04688, JP20H01858, JP18H01169 and JP18H01161) given by the Ministry of Education, Culture, Sports, Science and Technology of Japan (MEXT), and the Core-to-Core Program for Advanced Research Networks given by the Japan Society for the promotion of Science (JSPS). D.W. acknowledges support from the QUANOMET initiative (NL-4) and the Institute for Basic Science (Grant No. IBS-R009-Y3). P.L. acknowledges support by DFG within DFG Le967/16-1 and DFG EXC 2123 Quantum Frontiers.

1) L. Balents, Nature, **464**, 199 (2010).
2) A. Kitaev, Ann. Phys., **321**, 2 (2006).
3) Y. Motome, and J. Nasu, J. Phys. Soc. Japan, **89**, 012002 (2020).
4) J. Nasu, M. Udagawa, and Y. Motome, Phys. Rev. B, **92**, 115122 (2015).
5) B. J. Kim, H. Jin, S. J. Moon, J. Y. Kim, B. G. Park, C. S. Leem, J. Yu, T. W. Noh, C. Kim, S. J. Oh, J. H. Park, V. Durairaj, G. Cao, and E. Rotenberg, Phys. Rev. Lett., **101**, 076402 (2008).
6) G. Jackeli, and G. Khaliullin, Phys. Rev. Lett., **102**, 017205 (2009).
7) H. Liu, and G. Khaliullin, Phys. Rev. B, **97**, 014407 (2018).
8) R. Sano, Y. Kato, and Y. Motome, Phys. Rev. B, **97**, 014408 (2018).
9) S. H. Jang, R. Sano, Y. Kato, and Y. Motome, Phys. Rev. B, **99**, 241106 (2019).
10) Y. Singh, and P. Gegenwart, Phys. Rev. B, **82**, 064412

11) H. Kobayashi, M. Tabuchi, M. Shikano, H. Kageyama, and R. Kanno, J. Mater. Chem., **13**, 957 (2003).
12) M. Abramchuk, C. Ozsoy-Keskinbora, J. W. Krizan, K. R. Metz, D. C. Bell, and F. Tafti, J. Am. Chem. Soc., **139**, 15371 (2017).
13) K. W. Plumb, J. P. Clancy, L. J. Sandilands, V. V. Shankar, Y. F. Hu, K. S. Burch, H. Y. Kee, and Y. J. Kim, Phys. Rev. B, **90**, 041112 (2014).
14) K. Kitagawa, T. Takayama, Y. Matsumoto, A. Kato, R. Takano, Y. Kishimoto, S. Bette, R. Dinnebier, G. Jackeli, and H. Takagi, Nature, **554**, 341 (2018).
15) J. Q. Yan, S. Okamoto, Y. Wu, Q. Zheng, H. D. Zhou, H. B. Cao, and M. A. McGuire, Phys. Rev. Mater., **3**, 074405 (2019).
16) W. Yao, and Y. Li, Phys. Rev. B, **101**, 085120 (2020).
17) R. Zhong, T. Gao, N. P. Ong, and R. J. Cava, Sci. Adv., **6**, eaay6953 (2020).
18) J. Xing, E. Feng, Y. Liu, E. Emmanouilidou, C. Hu, J. Liu, D. Graf, A. P. Ramirez, G. Chen, H. Cao, and N. Ni, Phys. Rev. B, **102**, 014427 (2019).
19) Y. Hinatsu, and Y. Doi, J. Alloys Compd., **418**, 155 (2006).
20) J. G. Rau, E. K. H. Lee, and H. Y. Kee, Phys. Rev. Lett., **112**, 077204 (2014).
21) J. Chaloupka, and G. Khaliullin, Phys. Rev. B, **92**, 024413 (2015).
22) L. Janssen, E. C. Andrade, and M. Vojta, Phys. Rev. B, **96**, 064430 (2017).
23) S. M. Winter, Y. Li, H. O. Jeschke, and R. Valentí, Phys. Rev. B, **93**, 214431 (2016).
24) V. M. Katukuri, S. Nishimoto, V. Yushankhai, A. Stoyanova, H. Kandpal, S. Choi, R. Coldea, I. Rousochatzakis, L. Hozoi, and J. Van Den Brink, New J. Phys., **16**, 013056 (2014).
25) Y. Yamaji, Y. Nomura, M. Kurita, R. Arita, and M. Imada, Phys. Rev. Lett., **113**, 107201 (2014).
26) J. A. Sears, M. Songvilay, K. W. Plumb, J. P. Clancy, Y. Qiu, Y. Zhao, D. Parshall, and Y. J. Kim, Phys. Rev. B, **91**, 144420 (2015).
27) Y. Kobayashi, T. Okada, F. Ambe, K. Asai, M. Katada, and H. Sano, Inorg. Chem., **31**, 4570 (1992).
28) H. B. Cao, A. Banerjee, J. Q. Yan, C. A. Bridges, M. D. Lumsden, D. G. Mandrus, D. A. Tennant, B. C. Chakoumakos, and S. E. Nagler, Phys. Rev. B, **93**, 134423 (2016).
29) S.-Y. Park, S.-H. Do, K.-Y. Choi, D. Jang, T.-H. Jang, J. Schefer, C.-M. Wu, J. S. Gardner, J. M. S. Park, J.-H. Park, and S. Ji, arXiv:**1609**.05690.
30) B. Morosin, and A. Narath, J. Chem. Phys., **40**, 1958 (1964).
31) L. J. Sandilands, Y. Tian, K. W. Plumb, Y.-J. Kim, and K. S. Burch, Phys. Rev. Lett., **114**, 147201 (2015).
32) A. Banerjee, C. A. Bridges, J.-Q. Yan, A. A. Aczel, L. Li, M. B. Stone, G. E. Granroth, M. D. Lumsden, Y. Yiu, J. Knolle, S. Bhattacharjee, D. L. Kovrizhin, R. Moessner, D. A. Tennant, D. G. Mandrus, and S. E. Nagler, Nat. Mater., **15**, 733 (2016).
33) R. Yadav, N. A. Bogdanov, V. M. Katukuri, S. Nishimoto, J. Van Den Brink, and L. Hozoi, Sci. Rep., **6**, 37925 (2016).
34) Y. Kasahara, T. Ohnishi, Y. Mizukami, O. Tanaka, S. Ma, K. Sugii, N. Kurita, H. Tanaka, J. Nasu, Y. Motome, T. Shibauchi, and Y. Matsuda, Nature, **559**, 227 (2018).
35) R. D. Johnson, S. C. Williams, A. A. Haghighirad, J. Singleton, V. Zapf, P. Manuel, I. I. Mazin, Y. Li, H. O. Jeschke, R. Valentí, and R. Coldea, Phys. Rev. B, **92**, 235119 (2015).
36) N. I. Kolbin, I. N. Semenov, and Y. M. Shutov, Russ. J. Inorg. Chem., **8**, 1270 (1963).
37) M. A. McGuire, Q. Zheng, J. Yan, and B. C. Sales, Phys. Rev. B, **99**, 214402 (2019).
38) (Supplemental Material) The XRD patterns for $Os_xCl_3$ and $OsCl_4$ are provided online. The heat capacitis of $Os_xCl_3$ and $IrCl_3$ are also provided online.
39) T. Sakakibara, H. Mitamura, T. Tayama, and H. Amitsuka, Jpn. J. Appl. Phys., **33**, 5067 (1994).
40) Y. Luo, C. Cao, B. Si, Y. Li, J. Bao, H. Guo, X. Yang, C. Shen, C. Feng, J. Dai, G. Cao, and Z. A. Xu, Phys. Rev. B, **87**, 161121 (2013).
41) N. R. Davies, C. V. Topping, H. Jacobsen, A. J. Princep, F. K. K. Kirschner, M. C. Rahn, M. Bristow, J. G. Vale, I. Da Silva, P. J. Baker, C. J. Sahle, Y. F. Guo, D. Y. Yan, Y. G. Shi, S. J. Blundell, D. F. McMorrow, and A. T. Boothroyd, Phys. Rev. B, **99**, 174442 (2019).
42) T. Dey, A. Maljuk, D. V. Efremov, O. Kataeva, S. Gass, C. G. F. Blum, F. Steckel, D. Gruner, T. Ritschel, A. U. B. Wolter, J. Geck, C. Hess, K. Koepernik, J. Van Den Brink, S. Wurmehl, and B. Büchner, Phys. Rev. B, **93**, 014434 (2016).
43) A. Glamazda, P. Lemmens, S. H. Do, Y. S. Kwon, and K. Y. Choi, Phys. Rev. B, **95**, 174429 (2017).
44) G. Li, X. Chen, Y. Gan, F. Li, M. Yan, F. Ye, S. Pei, Y. Zhang, L. Wang, H. Su, J. Dai, Y. Chen, Y. Shi, X. Wang, L. Zhang, S. Wang, D. Yu, F. Ye, J. W. Mei, and M. Huang, Phys. Rev. Mater., **3**, 023601 (2019).
45) L. J. Sandilands, Y. Tian, A. A. Reijnders, H. S. Kim, K. W. Plumb, Y. J. Kim, H. Y. Kee, and K. S. Burch, Phys. Rev. B, **93**, 075144 (2016).
46) B. W. Lebert, S. Kim, V. Bisogni, I. Jarrige, A. M. Barbour, and Y. J. Kim, J. Phys. Condens. Matter, **32**, 144001 (2020).
47) T. H. Han, J. S. Helton, S. Chu, D. G. Nocera, J. A. Rodriguez-Rivera, C. Broholm, and Y. S. Lee, Nature, **492**, 406 (2012).
48) A. Glamazda, P. Lemmens, S.-H. Do, Y. S. Choi, and K.-Y. Choi, Nat. Commun., **7**, 12286 (2016).
49) J. Knolle, G. W. Chern, D. L. Kovrizhin, R. Moessner, and N. B. Perkins, Phys. Rev. Lett., **113**, 187201 (2014).
50) J. Nasu, J. Knolle, D. L. Kovrizhin, Y. Motome, and R. Moessner, Nat. Phys., **12**, 912 (2016).

51) Y. Singh, S. Manni, J. Reuther, T. Berlijn, R. Thomale, W. Ku, S. Trebst, and P. Gegenwart, Phys. Rev. Lett., **108**, 127203 (2012).
52) J. Chaloupka, and G. Khaliullin, Phys. Rev. B, **94**, 064435 (2016).
53) G. Cao, T. F. Qi, L. Li, J. Terzic, V. S. Cao, S. J. Yuan, M. Tovar, G. Murthy, and R. K. Kaul, Phys. Rev. B, **88**, 220414 (2013).
54) S. Manni, Y. Tokiwa, and P. Gegenwart, Phys. Rev. B, **89**, 241102 (2014).
55) S. H. Do, C. H. Lee, T. Kihara, Y. S. Choi, S. Yoon, K. Kim, H. Cheong, W. T. Chen, F. Chou, H. Nojiri, and K. Y. Choi, Phys. Rev. Lett., **124**, 47204 (2020).